# Analysis of Reactivity Induced Accident for Control Rods Ejection with Loss of Cooling

Hend Mohammed El Sayed Saad[1], Hesham Mohammed Mohammed Mansour[2] and Moustafa Aziz Abd El Wahab[1]

*1. Nuclear and Radiological Regulatory Authority, Nasr City, Cairo 11762, Egypt*

*2. Physics Department, Faculty of Science, Cairo University, Giza 12613, Egypt*

**Abstract:** Understanding of the time-dependent behavior of the neutron population in nuclear reactor in response to either a planned or unplanned change in the reactor conditions, is a great importance to the safe and reliable operation of the reactor. In the present work, the point kinetics equations are solved numerically using stiffness confinement method (SCM). The solution is applied to the kinetics equations in the presence of different types of reactivities and is compared with different analytical solutions. This method is also used to analyze reactivity induced accidents in two reactors. The first reactor is fueled by uranium and the second is fueled by plutonium. This analysis presents the effect of negative temperature feedback with the addition positive reactivity of control rods to overcome the occurrence of control rod ejection accident and damaging of the reactor. Both power and temperature pulse following the reactivity- initiated accidents are calculated. The results are compared with previous works and satisfactory agreement is found.

## 1. Introduction

Reactivity - initiated accident is nuclear reactor accident that involves inadvertent removal of control element from an operating reactor, thereby causing a rapid power excursion in the nearby fuel elements and temperature. The postulated scenarios for reactivity - initiated accidents are therefore focused on few events, which result in exceptionally large reactivity excursions, and therefore are critical to fuel integrity. In compared reference model [1], reactivity - initiated accident was considered to be due to negative temperature feedback. In the present work, we consider reactivity accident to be due to negative temperature feedback, and the addition positive reactivity of control rods to prevent such accidents of control rods ejection. We analyzed accidents in different types of reactors, e.g. [1] modular high temperature gas cooled reactor design like HTR-M and modular fast reactor design like PRISM, [1] using the stiffness confinement method for solving the kinetic equations. The stiffness confinement method (SCM) is used to solve the kinetics equations and overcome the stiffness problem in reactor kinetics [2]. The idea is based on the observation of stiffness characteristic, which is present only in the time response of the prompt neutron density, but not in the delayed neutron precursors. The method is therefore devised to have the stiffness decoupled from the differential equation for precursors and is confined to the one for prompt neutrons, which can be solved [2]. Numerical examples of applying the method to variety problems are given. The method is also used to analyze the reactivity induced accidents in two reactors data, modular high temperature gas cooled reactor (HTR-M) which is fueled by uranium and modular fast reactor design (PRISM) which is fueled by plutonium. [1] In the next, we discuss the mathematical method; present the results and discussion, and give the conclusion.

## 2. Experiments

The stiffness confinement method is used to overcome the stiffness problem in reactor kinetics for solving the point kinetics equations. The point kinetics equations are system of coupled ordinary differential equations, whose solution give the neutron density and delayed neutron precursor concentrations in tightly coupled reactor as a function of time. Typically these equations are solved using reactor model with at least six delayed precursor groups, resulting in system consisting of seven coupled differential equations. Obtaining accurate results is often problematic because the equations are stiff with many techniques, where very small time steps are used. These equations take the following form with an arbitrary reactivity function [3, 4]:

$$\frac{dn(t)}{dt} = \frac{\rho(t)-\beta}{\Lambda} n(t) + \sum_{i=1}^{6} \lambda_i C_i(t) \quad (1)$$

$$\frac{dC_i(t)}{dt} = \frac{\beta_i}{\Lambda} n(t) - \lambda_i C_i(t) \quad (2)$$

where: $n(t)$ is the time-dependent neutron density, or (power or neutron flux) all units are (MW) as power unit; $C_i(t)$ is the $i^{th}$ group delayed neutron precursor concentration or delayed neutron emitter population or precursor density ("latent-neutron" density or latent power; same units as in the power); $i$ is the number of precursor group; $\rho(t)$ is the time-dependent reactivity; $\beta_i$ is $i^{th}$ group delayed neutron fraction, and $\beta = \Sigma_i \beta_i$, is the total delayed neutron fraction. In addition, $\Lambda$ is the neutron generation time (s) and $\lambda_i$ is decay constant of the $i^{th}$-group delayed neutron emitters ($s^{-1}$). Introducing a set of "Reduced" precursor density functions $\hat{C}_i(t)$ and neutron density, through the following equation [2]:

$$C_i(t) = \hat{C}_i(t) \exp[\int_0^t u(t') dt'] \quad (3)$$

and defining two auxiliary functions $w(t)$ and $u(t)$, as in Eqs. (4) and (5):

$$w(t) = \frac{d}{dt} \ln n(t) \quad (4)$$

The function $w(t)$ is defined in the same way as Eq. (9) below and provides the mechanism key of the SCM. The function $u(t)$, however, has nothing to do with stiffness decoupling and is not really required theoretically. Since an exponential behavior is often characteristic for the first, order differential equations, however, a proper choice of $u(t)$ may make $\hat{C}_i(t)$ vary more slowly in time and thus expedite the numerical calculation. Choose the following $u(t)$ [2]:

$$u(t) = \frac{d}{dt} \ln S(t) \quad (5)$$

Where, $S(t)$ is defined by Eq. (7) as the sum over all $\lambda_i \cdot C_i(t)$. We can rewrite Eqs. (1) and (2) as follows [2]:

$$\frac{d\hat{C}_i(t)}{dt} = \left[ \frac{\beta_i}{\Lambda w(t) + \beta - \rho(t)} \right] \sum_{i=1}^{6} \lambda_i C_i(t) - [u(t) + \lambda_i] \hat{C}_i(t) \quad (6)$$

$$S(t) = \left[ \sum_{i=1}^{6} \lambda_i \hat{C}_i(t) \right] \exp\left[ \int_0^t u(t') dt' \right] \quad (7)$$

and

$$\frac{dn(t)}{dt} = \frac{\rho(t)-\beta}{\Lambda} n(t) + S(t)$$

Suppose that it is always possible to express:

$$n(t) = \exp[\int_0^t w(t') dt'] \quad (8)$$

and rewrite Eq. (1) as:

$$n(t) = \frac{\sum_{i=1}^{6} \lambda_i C_i(t)}{\left(w(t) + \left[\frac{\beta - \rho(t)}{\Lambda}\right]\right)} \qquad (9)$$

Eqs. (6)-(9) form the complete set of kinetic equations for the SCM. The initial conditions to be satisfied are:

$$u(0) = 0 \qquad (10a)$$

$$w(0) = \frac{\rho(0)}{\Lambda} \qquad (10b)$$

$$n(0) = n_0 \qquad (10c)$$

and

$$\hat{C}_i(0) = \frac{n_0 \beta_i}{\Lambda \lambda_i} \qquad (10d)$$

By using the initial conditions, we can obtain the numerical solution of the equations. We first start by setting $w$ and $u$ in Eq. (7) at their initial values and solves Eq. (7) for $\hat{C}_i$ by discretizing the equation in $t$. Having obtained $\hat{C}_i$, we calculate $S(t)$ with Eq. (1). Then, we use Eq. (5) to re-evaluate $w(t)$, plug it back into Eq. (7), and repeat the process until $w$ converges (requiring 50 iterations). Calculation for the current time step is then finished with an evaluation of the output value of $w$ and $u$ via Eqs. (5) and (10). Afterward, we predict the input values of $w$ and $u$ for the next time step by linear extrapolation from their output values in the previous and current time steps, and repeat the whole process of calculation for the next time step. It should be emphasized that within each time step, there is iteration to convergence on $w$ but no iteration for the function $u$, because $u$ is not required by the theory of (SCM) and is, in principle, with an arbitrary independent function chosen only to expedite the computation. Computer program is designed with programming languages (FORTRAN and MATLAB) codes to solve the above equations numerically using Runge-Kutta method, and the output power and temperature are determined under different input reactivities.

## 3. Model Problems

The SCM is tested with three types of problems which are:
(1) Step reactivity insertion,
(2) Ramp input,
(3) Sinusoidal input.

The results are compared against those obtained with other methods, e.g., Henry's $\theta$, weighted method [5], Exact data obtained with Ref. [2], and Taylor Series Methods [4, 6], CORE [7], Mathematica's built-in differential equation solver (implicit Runge-Kutta). Each of these methods is highly accurate, but they vary widely in their complexity of implementation.

*3.1 Step Reactivity Insertion*

Considering a kinetic problem with step reactivity insertion with $\beta = 0.007$. In this case, $\rho(t) = \rho_0$ for $t \geq 0$. The following input parameters were used: $\lambda_i$ (s$^{-1}$) = (0.0127, 0.0317, 0.155, 0.311, 1.4, 3.87), $\beta_i$ = (0.000266, 0.001491, 0.001316, 0.002849, 0.000896, 0.000182) and $\Lambda$ = 0.00002 s. Four step reactivity insertions are considered: two prompt subcritical $\rho$ = 0.003 and 0.0055, one prompt critical $\rho$ = 0.007, one prompt supercritical $\rho$ = 0.008 [2, 7]. The values of $n(t)$ obtained with the present work are compared (Table 1) with those obtained with a code based on the so-called "Henry's θ, weighted method", which modifies finite difference equations by introducing tactically chosen weighting functions. The step size taken was $h_1 = 0.001$. For comparison, we chose "Henry's θ, weighting method", and the exact values that obtained from Ref. [2] with the present results. The numbers presented in Table 1 are computed with time steps (1 s, 10 s and 20 s). The results indicate that the present model solutions are in good agreement with all results. The iteration in computing was used for repeating the process until $w$ and $u$ converge (requiring approximately 100 iterations) to get step reactivity insertion with accurate results which are compared with several methods.

*3.2 Ramp Input of Reactivity*

Consider now the two cases of ramp input. Ramp reactivity usually takes the form:
$$\rho(t) = \rho_0 \, t$$

Where, $\rho_0 = \dfrac{\rho}{\beta}$

Is a given reactivity expressed in dollars [10, 11]. We will use the same parameters, which are used in the step reactivity example, and compare our results with those of Ref. [2]. The first case is extremely fast and the second is moderately fast. In the first one, it can be seen that, the response of reactor core at 0.001 s after a ramp input of reactivity at the rate of $100/s is calculated (with six groups of delayed neutron). The computational results for this case are presented in Table 2 in comparison with the SCM solution by Ref. [2].

**Table 1  Comparison of present work and different methods for step reactivity insertion.**

| ρ | Method | n(t) | | |
|---|---|---|---|---|
| | | $t = 1$ s | $t = 10$ s | $t = 20$ s |
| 0.003 | Present results | 2.1849 | 7.89116 | 27.8266 |
| | θ-weighting | 2.1737 | 8.0069 | 28.076 |
| | SCM | 2.2254 | 8.0324 | 28.351 |
| | Exact | 2.2098 | 8.0192 | 28.297 |
| | | $t = 0.1$ s | $t = 2$ s | $t = 10$ s |
| 0.0055 | Present results | 5.16136 | 42.5859 | 1.37302 + 05 |
| | θ-weighting | 5.19450 | 42.6520 | 1.38820 + 05 |
| | SCM | 5.20570 | 43.0240 | 1.38750 + 05 |
| | Exact | 5.21000 | 43.0250 | 1.38860 + 05 |
| | | $t = 0.01$ s | $t = 0.5$ s | $t = 2$ s |
| 0.007 | Present results | 4.44702 | 53.0908 + 02 | 20.4510 + 10 |
| | θ-weighting | 4.50891 | 53.4840 + 02 | 20.6410 + 10 |
| | SCM | 4.50013 | 53.5302 + 02 | 20.6270 + 10 |
| | Exact | 4.50882 | 53.4593 + 02 | 20.5912 + 10 |
| | | $t = 0.01$ s | $t = 0.1$ s | $t = 1$ s |
| 0.008 | Present results | 6.14858 | 1.17679 + 03 | 6.0564 + 23 |
| | θ-weighting | 6.20300 | 1.41150 + 03 | 6.2258 + 23 |
| | SCM | 6.20460 | 1.40891 + 03 | 6.1574 + 23 |
| | Exact | 6.20291 | 1.41042 + 03 | 6.1634 + 23 |

The second case is a (moderately fast) ramp of $0.01/s to reactor core. The values of the physical parameters are the same as those of step reactivity insertion examples. The computational results for this case are presented in Table 3 along with other methods. The iteration in computing was used for repeating the process until $w$ and $u$

converge (requiring 10 iterations) to take ramp reactivity insertion which is a time dependent function with small time step in order to get accurate results in comparison with several methods.

*3.3 Sinusoidal Input of Reactivity*

Consider the case of sinusoidal reactivity. In this case the kinetic parameters are used: $\lambda_i$ (s$^{-1}$) = (0.0124, 0.0305, 0.111, 0.301, 1.14 and 3.01), $\beta_i$ = (0.000215, 0.001424, 0.001274, 0.002568, 0.000748, and 0.000273), $\Lambda$ = 0.0005 s, $T$ = 5.00 s and $\beta$ = 0.006502. The reactivity is a time dependent function of the form [2, 4, and 10]:

$$\rho(t) = \rho_0 \sin(\frac{\pi t}{T})$$

Where, $T$ is a half-period and $\rho_0 = \beta$. The results of the present method are compared with other methods in Table 4 and showed a good agreement. The iteration in the computation, is used for repeating the process until $w$ and $u$ converge (requiring 100 iterations) to get step reactivity insertion with accurate results. The iteration in computing is used for repeating the process until $w$ and $u$ converge (requiring 10 iterations) to take sinusoidal reactivity insertion which is a triangular function inside it half period and small time step to get accurate results. The results are compared with several methods.

**Table 2** Comparison of present work and SCM method in Ref. [2] for ramp input of reactivity the first case: ( extremely fast).

| $\rho$ | Methods | $n(t)$ |
|---|---|---|
| $\rho_0 = 0.7$ | Present results | 1.09643 |
| | SCM | 1.10842 |

**Table 3** Comparison of present work and SCM method for ramp input of reactivity the second case :( moderately fast).

| Methods | $t = 2$ s | $t = 4$ s | $t = 6$ s | $t = 8$ s | $t = 9$ s |
|---|---|---|---|---|---|
| Present results | 1.32081 | 2.19494 | 5.49151 | 4.20720 + 01 | 4.79378 + 02 |
| θ-Weighting | 1.33832 | 2.22903 | 5.58852 | 4.32151 + 01 | 5.06363 + 02 |
| SCM | 1.33824 | 2.22842 | 5.58191 | 4.27882 + 01 | 4.87814 + 02 |
| Exact | 1.33739 | 2.22832 | 5.58151 | 4.27811 + 01 | 4.87452 + 02 |

**Table 4** Comparison of present work and other methods for sinusoidal reactivity.

| Methods | $t = 2$ s | $t = 4$ s | $t = 6$ s | $t = 8$ s | $t = 10$ s |
|---|---|---|---|---|---|
| Present results | 11.320 | 84.950 | 14.4824 | 7.87237 | 12.1093 |
| Taylor | 11.3820 | 92.2761 | 16.0317 | 8.63622 | 13.1987 |
| Core | 10.1475 | 96.7084 | 16.9149 | 8.89641 | 13.1985 |
| Mathematica | 11.3738 | 92.5595 | 16.0748 | 8.65512 | 13.2202 |

## 4. Analysis of Reactivity Initiated Accident

*4.1 Reactivity Initiated Accident*

Reactivity - initiated accident involves an unwanted increase in fission rate and reactor power. Power increase may damage the reactor core, and in very severe cases, even lead to the disruption of the reactor. The immediate consequence of reactivity - initiated accident is fast rise in fuel power and temperature. The power excursion may lead to failure of the nuclear fuel rods and release radioactive material into primary reactor coolant. In this study, a new computer program has been developed for simulating the reactor dynamic behavior during reactivity induced transients, and it has been used for the analysis of specified reactivity - initiated accidents considered in several cases. We introduce the two models reactors with system parameters, which are characteristic for modular high temperature gas-cooled reactor design like HTR-M [8] and modular fast reactor design like PRISM [9]. For simplicity, we refer to these models of two reactors as HTR-M and PRISM (Tables 5 and 6). For delayed neutron parameters, it is assumed that, HTR-M is fuelled by $^{235}$U and PRISM by $^{239}$Pu as fissile nuclides. The dynamic equations for the two models are the conventional of the point reactor kinetics equations in combination with linear temperature feedback in reactivity, an adiabatic heating of the core after loss of cooling [1], where Eq. (13a) may be modified to add positive reactivity of control rods. The data of the two reactor models are given in Tables 5-7:

**Table 5** $^{235}$U (thermal neutrons).

| $\lambda_i$ (sec$^{-1}$) | 0.0124 | 0.0305 | 0.111 | 0.301 | 1.14 | 3.01 |
|---|---|---|---|---|---|---|
| $\beta_i$ | 0.000215 | 0.001424 | 0.001274 | 0.002568 | 0.0007485 | 0.0002814 |
| | $\beta_{tot}$ = 0.0067 | | | | $\Lambda$ = 1.00E-4 (s) | |

**Table 6** $^{239}$PU (fast neutrons).

| $\lambda_i$ (sec$^{-1}$) | 0.0129 | 0.0311 | 0.134 | 0.331 | 1.26 | 3.21 |
|---|---|---|---|---|---|---|
| $\beta_i$ | 7.6E-005 | 5.6E-004 | 4.32E-004 | 6.56E-004 | 2.06E-004 | 7.00E-005 |
| | $\beta_{tot}$ = 0.0020 | | | | $\Lambda$ = 1.00E-7 (s) | |

**Table 7** Adiabatic inherent shutdown data for two model reactors.

| Types of reactors | $n_0$ (MW) | $c$ (MJ/K) | $\alpha$ (K$^{-1}$) |
|---|---|---|---|
| HTR-M | 200.00 | 100.00 | 2.2E-005 |
| PRISM | 470.00 | 200.00 | 9.00E-006 |

$$\frac{dn(t)}{dt} = \frac{\rho_{net}(t) - \beta}{\Lambda} n(t) + \sum_{i=1}^{6} \lambda_i C_i(t) \quad (11)$$

$$\frac{dC_i(t)}{dt} = \frac{\beta_i}{\Lambda} n(t) - \lambda_i C_i(t) \quad (12)$$

$$\rho_{net}(t) = \rho_{feed}(t) + \rho_{ext}(t) \quad (13a)$$

$$\rho_{feed} = -\alpha(T(t) - T_0)$$

$$\rho_{ext} = \rho_{CR} = \rho_{cr1} \text{ or } \rho_{cr2} \text{ or } \rho_{cr3} \text{ or } \rho_{cr4} \quad (13b)$$

$\rho_{feed}$ = *feedback reactivity*

$\rho_{ext}$ = *external reactivity* = *control rods reactivity*

$$\frac{dT(t)}{dt} = \frac{1}{c} n(t) \quad (14)$$

where $n(t)$ = reactor power (MW); $\rho_{net}(t)$ = the time-dependent reactivity function; $\rho_{CR}$ = addition positive reactivity of control rods; $\beta$ = total delayed neutron fraction; $\beta = \Sigma_i \beta_i \cdot \beta_i$ = delayed neutron faction of $i^{th}$ group; $\Lambda$ = neutron generation time (s); $\lambda_i$ = decay constant of $i^{th}$ group delayed neutron emitters (s)$^{-1}$; $C_i(t)$ = delayed neutron emitter population (in power units); $\alpha$ = negative temperature coefficient of reactivity (K$^{-1}$); $T$ = reactor temperature (K); $T_0$ = critical reactor temperature (K) and c=heat capacity of reactor (MJ/K).

In the equation of total reactivity $\rho(t)$, the additional positive reactivity of control rods $\rho_{cr}$ has four cases to prevent the control rods ejection accident:

$$\rho_{cr1} = \rho_1 = 0, \rho_{cr2} = \rho_2 = (\beta/2),$$

$$\rho_{cr3} = \rho_3 = (0.8\beta), \rho_{cr4} = \rho_4 = (\beta).$$

The input parameters of the kinetic equations for two types of reactors with different fissile materials are shown in Tables 5 and 6.

*4.2 Reactivity Evaluation*

The reactivity of one, two and three control rods worth are calculated based on the assumptions of relating control rods worth by the delayed neutron fraction $\beta$. Assuming that, the ejection of one, two and three rods could induce positive reactivity as indicated in Table 8 for each type of reactors, in the two models.

**Table 8  Additional positive reactivity of control rods insertion.**

| No. of control rods | $\rho$ (in \$) for $U^{235}$ | $\rho$ (in \$) for $PU^{239}$ |
|---|---|---|
| 1 | 0.5 | 0.5 |
| 2 | 0.8 | 0.8 |
| 3 | 1.00 | 1.00 |

Section (II)

## 5. Results and Discussion

In this section, we explain the result of the total energy production which is expressed in full-power-seconds (FPS) by dividing the energy through nominal power, asymptotic temperature increase, and equilibrium reactivity after shutdown. The relevant quantity in relation to reactor safety is the asymptotic temperature increase as determined by the total energy production during autonomous shutdown and by the heat capacity. There is simple relation between energy produced and temperature increase due to the absence of heat loss [1]. As proved in Ref. [1]:

$$E_\infty^2 = \frac{2\Lambda C}{\alpha} * n_0 \left\{ 1 + \frac{1}{\Lambda} \sum_i \frac{\beta_i}{\lambda_i} \right\} \approx \frac{2n_0 C}{\alpha} \sum_i \frac{\beta_i}{\lambda_i} \quad (15)$$

Where, $E_\infty$ = total fission energy produced during autonomous shutdown; $n_0$ = the initial reactor power condition depending on the type of reactor. In case of autonomous shutdown, which use Eq. (13a) and Eq. (14) as in Ref. [1], that the asymptotic temperature $T_\infty$ and equilibrium reactivity after shutdown $\rho_\infty$ are found in Eq. (16) as:

$$T_\infty - T_0 = [\frac{2n_0}{\alpha\, c} \sum_i \frac{\beta_i}{\lambda_i}]^{1/2}, \quad \rho_\infty = -[\frac{2\alpha n_0}{c} \sum_i \frac{\beta_i}{\lambda_i}]^{1/2} \quad (16)$$

The total energy production quantity is physically more appealing; it can be seen that, both reactors, the fission energy production during adiabatic inherent shutdown is equivalent to one minute full-power operation. As a consequence of the high heat capacity of the cores, this energy is easily accommodated with temperature increase about 130 K. So, the results of the two reactors models are given in Table 9.

## 6. Reactivity Addition at Full Power Condition

*6. 1 First Reactor (PRISM Reactor)*

PRISM reactor is assumed to be operating at equilibrium power condition equal to 470 (MW) and the limited value of time (s) on x axis equals to 300 (s) at full power condition. Where, control rod insertion increases the thermalization of neutrons, and thus, results in a positive reactivity addition. Control rod insertion requires a certain driving force. The driving forces on control rods in the reactors are the buoyancy from the fuel material and the supporting force from the control system of reactor. If the control system should lose the support of control rods or control rods should break, control rods would be flown out of the reactor. Thus, in PRISM reactor, accidental insertions can result from the ejection of control rod drive, and/or control rod control system or operator error. Reactivity is also added step by step. The full power transients for one, two, and three control rods ejection are shown in Fig. 1. When control rods are ejected, power pulse indicates in the four cases: First, with negative temperature feedback and without positive reactivity of control rods at initial condition $t = 0$ (s), $n_0 = 470$ (MW), power decreases and after $t = 300$ (s), power approaches to saturation with $n = 10$ (MW). Second, with negative temperature feedback and addition control rod reactivity equal ($\beta/2$), the maximum power ratio increase by 2.00 times from the initial value of power at $t = 0.000768$ (s). Third, with negative temperature feedback and addition control rod reactivity equal ($0.8\beta$), the maximum power ratio increase by 4.6149 times from the initial value of power at $t = 0.000582$ (s). Fourth, with negative temperature feedback and addition control rod reactivity equal ($\beta$), the maximum power ratio increase by 83.8085 times from the initial value of power at $t = 0.006160$ (s). When control rods are ejected, power pulse is increased many times than; the rated power is generated in a very short time. This is because the accident is reactivity accident.

The temperature transients are shown in Fig. 2 for four cases. Because of temperature is proportional with power at Eq. (14) and with Eq. (13a) with the net reactivity, so that, power in Eq. (11) increases due to the positive reactivity addition in reactivity equation of control rod worth. The maximum temperature exceeded 1,521 K for about 250 s. The results indicate that, in the first case at initial condition $t = 0$ (s), $T_0 = 950$ (K), after that temperature increases until $t = 250$ s become at first case: $T = 1,195$ (K), second case: $T = 1,330$ (K), third case: $T = 1,430$ (K), fourth case: $T = 1,503$ (K). After $t = 200$ (s), temperature approaches to saturation.

**Table 9** Adiabatic inherent shutdown results for two models reactors.

| Types of reactors | | $E_\infty/n_0$ (fps) | $T_\infty - T_0$ (K) | $\rho_\infty$ (%) |
|---|---|---|---|---|
| H. Van Dam | HTR-M | 62 | 124 | -0.27 |
| Present Results | | 63 | 126 | -0.272 |
| H. Van Dam | PRISM | 55 | 129 | -0.12 |
| Present results | | 53 | 123 | -0.117 |

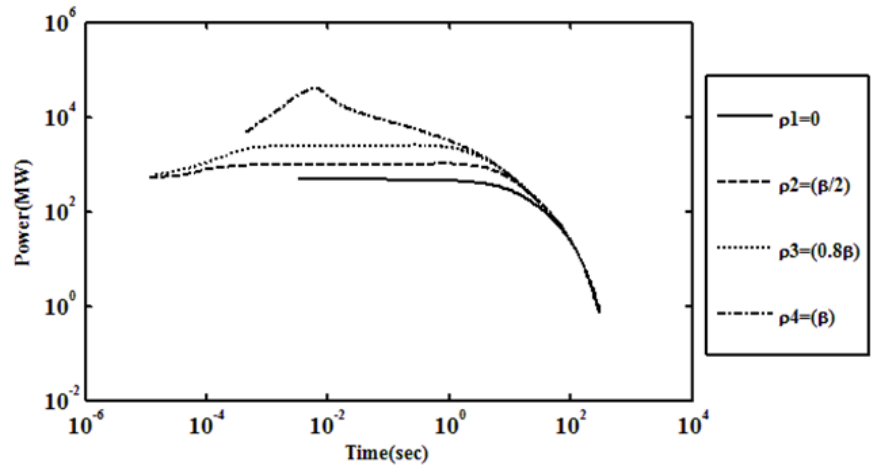

Fig. 1   The power (MW) transient as a function of time at full power condition with different values of positive of control rods ejection for PRISM reactor.

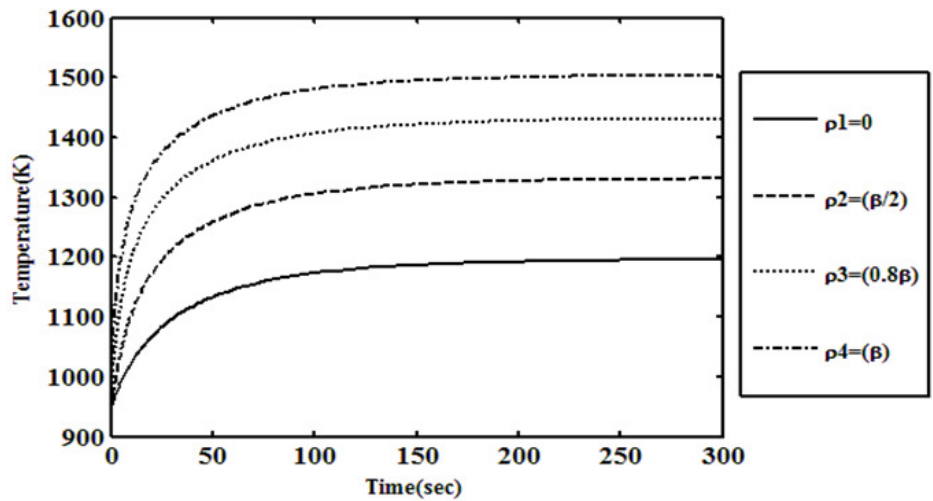

Fig. 2   The temperature (K) as a function of time during the transients at full power for PRISM reactor.

*6.2 Second reactor (HTR-M Reactor)*

HTR-M reactor is assumed to be operating at equilibrium power condition equal to 200 (MW) and the limited value of time (s) on x axis equals to 300 (s) at full power condition. Reactivity is also added step by step as, explained above in PRISM reactor. The full power transients for one, two, and three control rods ejection are shown in Fig. 3. Control rods are ejected, power pulse indicates in four cases: First, with negative temperature feedback and without positive reactivity of control rods at initial condition $t = 0$ (s), $n_0 = 200$ (MW), power decreases due to negative temperature. Second, with negative temperature feedback and addition control rod reactivity equal ($\beta/2$), the maximum power ratio increase by 4.4585 times from the initial value of power at $t = 9.355$ (s). Third, with negative temperature feedback and addition control rod reactivity equal ($0.8\beta$), the maximum power ratio increase by 15.3450 times from the initial value of power at $t = 2.783$ (s). Fourth, with negative temperature feedback and addition control rod reactivity equal ($\beta$), power ratio increases by 53.50 times from the initial value of power at $t = 0.706$ (s). When control rods are ejected, power pulse is increased many times of the rated power is generated in a very short time. This is because the accident is a reactivity accident.

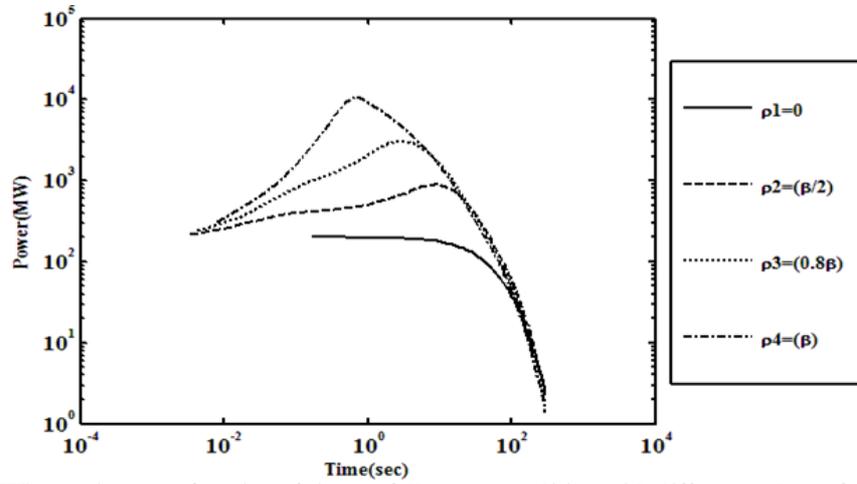

**Fig. 3  The power (MW) transient as a function of time at full power condition with different values of positive reactivity of control rods ejection for HTR-M reactor.**

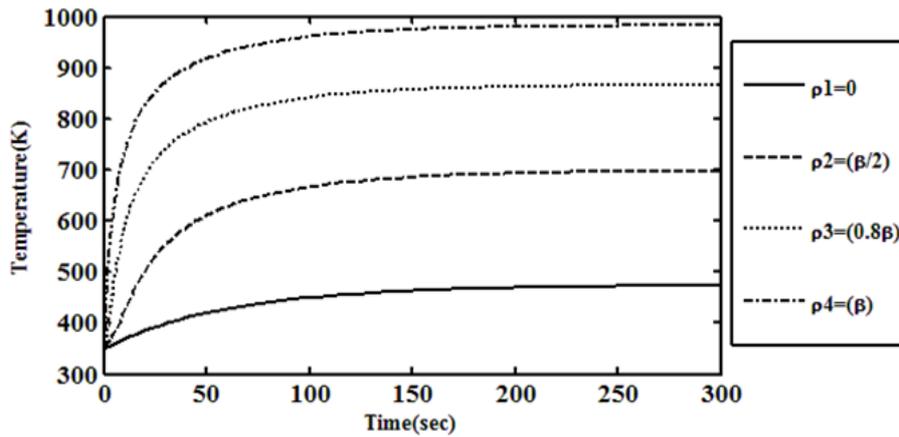

**Fig. 4  The temperature (K) as a function of time during the transients at full power for HTR-M reactor.**

The temperature transients are shown in Fig. (4) for four cases. Because of temperature is proportional with power at Eq. (14) and with Eq. (13a), with the net reactivity, so that power in Eq. (11) increases due to the positive reactivity addition in reactivity equation of control rod worth. The maximum temperature exceeded 998.4K for about 250 s. The results indicate that, in the first the initial condition $t = 0$ (s), $T_0 = 350$ (K), after that temperature increase until $t = 250$ (s) become at first case: $T = 476.4$ (K), second case: $T = 696.7$ (K), third case: $T = 865.6$ (K), fourth case: $T = 982.5$ (K). After $t = 200$ (s), temperature approaches to saturation.

## 7. Conclusions

Computer program is designed to solve the point reactor dynamics equations using the stiffness confinement method (SCM) and different input reactivity is applied (step, ramp and sinusoidal), the resultant powers are determined and illustrated. Good accuracy in comparison with reference values is obtained.

The model is applied to the two types of reactors. There are modular of fast reactor design like PRISM reactor [9] and modular high temperature gas-cooled reactor design like HTR-M reactor [8]. PRISM reactor is fuelled by $^{239}$Pu, the HTR-M reactor is fuelled by $^{235}$U as fissile nuclides.

In the work of Van Dam [1] (we used it for comparison purpose), the author obtained reactivity accident due to negative temperature feedback after loss of cooling to different reactors with different fissile material. Reactivity - initiated accident is considered to be due to linear temperature feedback and an adiabatic heating of the core after loss of cooling. In the present work, we consider reactivity accident due to linear temperature feedback, an adiabatic heating of the core after loss of cooling and with addition of positive reactivity due to control rods ejection.

We analyzed accidents in different types of reactors (HTR-M and PRISM), using the stiffness confinement method for solving the kinetics equations. In the present work, one obtains reactivity induced accident due to control rods ejection with negative temperature feedback and addition of positive reactivity of the control rods to overcome the occurrence of control rods ejection accident and prevent reactors from damage. The addition of positive reactivity is used for four cases: (0, $\beta$/2, 0.8$\beta$, $\beta$), where at the zero case only negative temperature feedback as the case of the Ref. [1] and the other cases of negative temperature feedback and the addition value of control rods reactivity. This is called reactivity induced accident.

The power for $^{239}$Pu fueled reactor, when reactivity of reactor is increased by $\beta$, the reactor peak power increased by 83. 8,085 times the initial value with the saturated temperature of 1,503 (K).

For HTR-M reactor increase by factor of 53.5 times the initial value at equilibrium temperature of 1,000 (K), when reactivity is increased by $\beta$.

**References**

[1] H. Van Dam, Dynamics of passive reactor shutdown, Prog. Nucl. Energy 30 (1996) 255.
[2] Y. Chao, Al. Attard, A resolution to the stiffness problem of reactor kinetics, Nuclear Science and Engineering 90 (1985) 40-46.
[3] J.J. Duderstadt, L.J. Hamilton, Nuclear Reactor Analysis. John Wiley & Sons, 1976, pp. 233-251.
[4] D. McMahon, A. Pierson, A Taylor series solution of the reactor point kinetics equations, arXiv: 1001.4100 2 (2010) 1-13.
[5] T.A. Porsching, The numerical solution of the reactor kinetics equations by difference analogs: A comparison of methods, WAPD-TM-564, U.S. National Bureau of Standards, U.S. Department of Commerce (1966) 1-44.
[6] B. Mitchell, Taylor series methods for the solution of the point reactor kinetic equations, Annals of Nuclear Energy 4 (1977) 169-176.
[7] B. Quintero-Leyva, CORE: A numerical algorithm to solve the point kinetics equations, Annals of Nuclear Energy 35 (2008) 2136-2138.
[8] K. Kugeler, R. Schulten, High Temperature Reactor Technology, Springer, Berlin, 1989, pp. 246-260.
[9] G.J. Van Tuyle, G.C. Slovik, R.J. Kennett, B.C. Chan, A.L. Aronson, Analyses of unscrammed events postulated for the PRISM design, Nuclear Technology, 91 (1990) 165-184.
[10] M. Kinard, E.J. Allen, Efficient numerical solution of the point kinetics equations in nuclear reactor dynamics, Annals of Nuclear Energy 31 (2004) 1039-1051.
[11] D.L. Hetrick, Dynamics of Nuclear Reactors, University of Chicago Press, Chicago, 1971.